\documentclass[aps,pre,floatfix,twocolumn]{revtex4}
\usepackage{amssymb,epsf}
\begin{document}
\begin{widetext}
\noindent
{\fontfamily{cmss}\fontseries{sbc}\fontsize{22}{1}\selectfont
Scale-free behavior of the Internet global performance\\}

\noindent
{\fontfamily{cmss}\fontseries{bx}\fontsize{9}{9}\selectfont
Roberto Percacci$^*$ and Alessandro Vespignani$^{\dag}$\\}

\noindent
{\sf\fontsize{8}{0}\selectfont  
$^*$International School for Advanced Studies SISSA/ISAS,
via Beirut 4, 34014 Trieste, Italy; and 
$^{\dag}$The Abdus Salam International Centre for Theoretical Physics
  (ICTP), P.O. Box 586, 34100 Trieste, Italy\\}

\noindent
{\sf\fontsize{8}{0}\selectfont June 11, 2002.}
\thispagestyle{empty}
\end{widetext}

{\fontfamily{cmss}\fontseries{bx}\fontsize{9}{9}\selectfont
\noindent
 
Measurements and data analysis have proved very effective in the 
study of the Internet's physical fabric and have shown heterogeneities 
and statistical fluctuations extending over several orders of magnitude.
Here we analyze performance measurements obtained by the 
PingER monitoring infrastructure. We focus on the relationship between 
the Round-Trip-Time (RTT) and the geographical distance. 
We define  dimensionless variables that contain information on 
the quality of  Internet connections finding  
that their probability distributions are characterized by 
a slow power-law decay signalling the presence of scale-free features.
These results point out the extreme heterogeneity of the Internet 
since the transmission speed between different points of the network 
exhibits very large fluctuations.
The associated scaling exponents appear to have fairly stable
values in different data sets and thus define an 
invariant characteristic of the Internet that might be used in the future 
as a benchmark of the overall state of ``health'' of the Internet.
The observed scale-free character 
should be incorporated in models and analysis of Internet performance.\\
}

\fontfamily{cmr}\fontsize{9}{10}\selectfont
\noindent
The Internet is a self-organizing system whose size has 
already scaled five orders of magnitude since its inception.
Given the extremely complex and interwoven structure of the Internet, 
several research groups started to deploy
technologies and infrastructures aiming to obtain a more global
picture of the Internet. This has led to very interesting findings
concerning the Internet maps topology.
Connectivity and other metrics  are characterized by algebraic 
statistical distributions that signal fluctuations extending over 
many length scales \cite{faloutsos,gov00,caida,pvv,willinger}. 
These scale-free properties and the associated heterogeneity of 
the Internet fabric define a large scale object whose 
properties cannot be inferred from local ones, and are 
in sharp contrast with standard graph models.
The importance of a correct topological characterization of the 
Internet in routing protocols and the  parallel  
advancement in the understanding of scale-free networks \cite{barab99}
have triggered a renewed interest in Internet measurements and modeling.  
Considerable efforts have been devoted also to the collection of
end-to-end performance data by means of active measurements techniques.
This activity has stimulated several  studies that, however,  
focus mainly on individual properties of hosts, routers or routes.
Only recently, an increasing body of work focuses on the performance 
of the Internet as a whole, especially to forecast future performance 
trends \cite{paxson,leestep}. These measurements pointed out the presence
of highly heterogeneous performances and 
it is our interest to inspect the possibility of a cooperative ``emergent
phenomenon'' with associated scale-free behavior. 

The basic testing package for Internet performance is the original 
PING (Packet InterNet Groper) program. Based on the Internet Control 
Message Protocol (ICMP), Ping works much like a sonar echo-location, 
sending packets that elicit a reply from the targeted host. 
The program then measures the round-trip-time (RTT), i.e. how long 
it takes each packet to make the round trip. 
Organizations such as the National Laboratory for Applied Network 
Research (http://moat.nlanr.net/) and the Cooperative Association for 
Internet Data Analysis (http://www.caida.org/) use PING-like 
probes from geographically diverse monitors to collect RTT data 
to hundreds or thousands of Internet destinations.
Our Internetwork Performance Measurement (IPM) project  
currently participates in the PingER monitoring 
infrastructure (http://www-iepm.slac.stanford.edu/).      
PingER was developed by the Internet End-to-end Performance Measurement
(IEPM) group to monitor the end-to-end performance of Internet links.
It consists of a number of beacon sites sending regularly ICMP probes to
hundreds of targets and storing all data centrally.
Most beacons and targets are hosts belonging to universities or research
centers; they are connected to many different networks and backbones
and have a very wide geographical distribution, so they likely represent
a statistically significant sample of the Internet as a whole.

We have analyzed two years worth of PingER data, 
going from April 2000 to March 2002.
We have selected 3353 different beacon-target pairs, taken out of 36
beacons and 196 targets. For each pair we have considered the 
following metrics: the geographic distance of the hosts $d$ (measured
on a great circle), the monthly average packet loss rate $r$ (the
percentage of ICMP packet that does not reach the target point), 
the monthly minimum and  average round-trip-times RTT$_{min}$, 
and RTT$_{av}$, respectively. These data offer the opportunity to test
various hypotheses on the statistical behavior of Internet performance.
Each data point is the monthly summary of approximately 1450 single 
measurements.
The geographic position of hosts is known with great accuracy for some sites, 
but in most cases it may be wrong by 10-20km. 
Consequently, we have discarded pairs of sites that are less than 
this distance apart.
The end-to-end delay is governed by several factors. 
First, digital information travels along fiber optic cables 
at almost exactly 2/3 the speed of light in vacuum. 
This gives the mnemonically very convenient value of 1ms RTT per 100km
of cable. 
Using this speed one can express the geographic distance $d$ 
in light-milliseconds, obtaining an 
absolute physical lower bound on the RTT between sites.
The actual measured RTT is (usually) larger than this value because of 
several factors. 
First, data packets often follow rather circuitous paths 
leading them through a number of nodes that are far from the
geodesic line between the endpoints.
Furthermore, each link in a given path is itself far from being 
straight, often following highways, railways or power lines \cite{henk}.
The combination of these factors produces
a purely geometrical enhancement factor of the RTT.
In addition, there is a minimum processing delay $\delta$ 
introduced by each router along the way, of the order of 
50-250$\mu$s per hop on average, summing up to a few ms for a 
typical path \cite{henk}.
This can be significant for very close site pairs, but is negligible
for most of the paths in the PingER sample.
On top of this, the presence of cross traffic along the route can cause
data packets to be queued in the routers. Let $t_R$ be the sum 
of all processing and queueing delays due to the routers on a path.
When the traffic reaches congestion, $t_R$ becomes a very significant 
part of the RTT and packet loss also sets in.
We have considered minimum and average values of the RTT over 
one month periods. It is plausible that even on rather congested 
links there will be a moment in the course of a month when 
$t_R$ is negligible, so RTT$_{min}$ can be taken as 
an estimate of the best possible communication performance
on the given data path, subject only to the intrinsic geometrical 
enhancement factor and the minimum  processing delay. 
On the other hand, RTT$_{av}$ for a given site pair is obtained by considering 
the average RTT over one month periods. This takes into account also 
the average queueing delay and gives an estimate of the overall
communication performance on the given data path.
\begin{figure}
\centerline{\epsfxsize=7cm \epsfbox{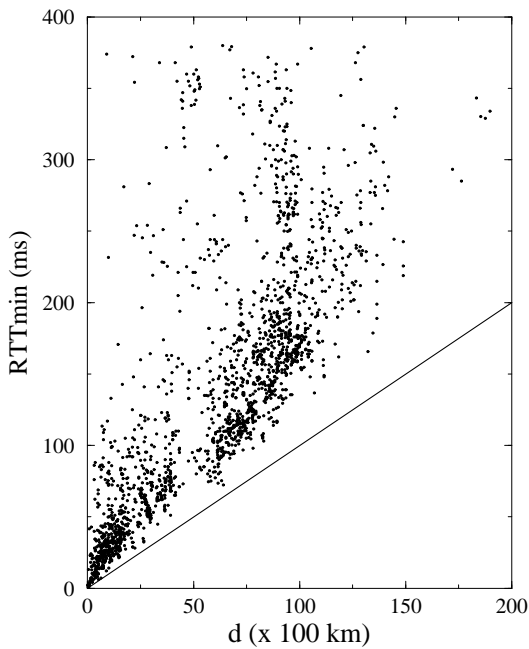}}
\caption{{\fontfamily{cmss}\fontseries{bx}\selectfont RTT$_{min}$ 
between 2114 host pairs} (PingER data set of 
February 2002) as a function of their distance $d$. Each point 
correspond to a different 
host pair. The line indicates the physical lower bound provided by the
speed of light in transmission cables. It is possible to observe the very 
large fluctuations in the RTT$_{min}$ of different host pairs 
separated by the same distance. For graphical reasons the picture frame 
is limited to 400ms, however, several outliers up to 900ms are present 
in the data set.}
\label{fig1}
\end{figure}

We studied the level of correlation between geographic distance and the 
RTT$_{min}$ and RTT$_{av}$ of source-destination pair. 
In Fig.\ref{fig1} we
report the obtained relationship for RTT$_{min}$  compared 
with the solid line representing the speed of light in optic fibers 
at each distance.  
While it is possible to observe a linear correlation of the 
RTT$_{min}$ with the physical distance of hosts, yet the data 
are extremely scattered. The RTT$_{av}$ present a qualitatively 
very similar behavior, and it is worth remarking that both 
plots are  in good agreement with similar analysis obtained for 
different data sets \cite{leestep,huff1,huff2}. 
While several qualitative features of this plot provide insight into
the geographical distribution of hosts and their connectivity, 
it misses a quantitative characterization of the intrinsic fluctuations
of performances and their statistical properties. 

\begin{figure}
\centerline{\epsfxsize=8.0cm \epsfbox{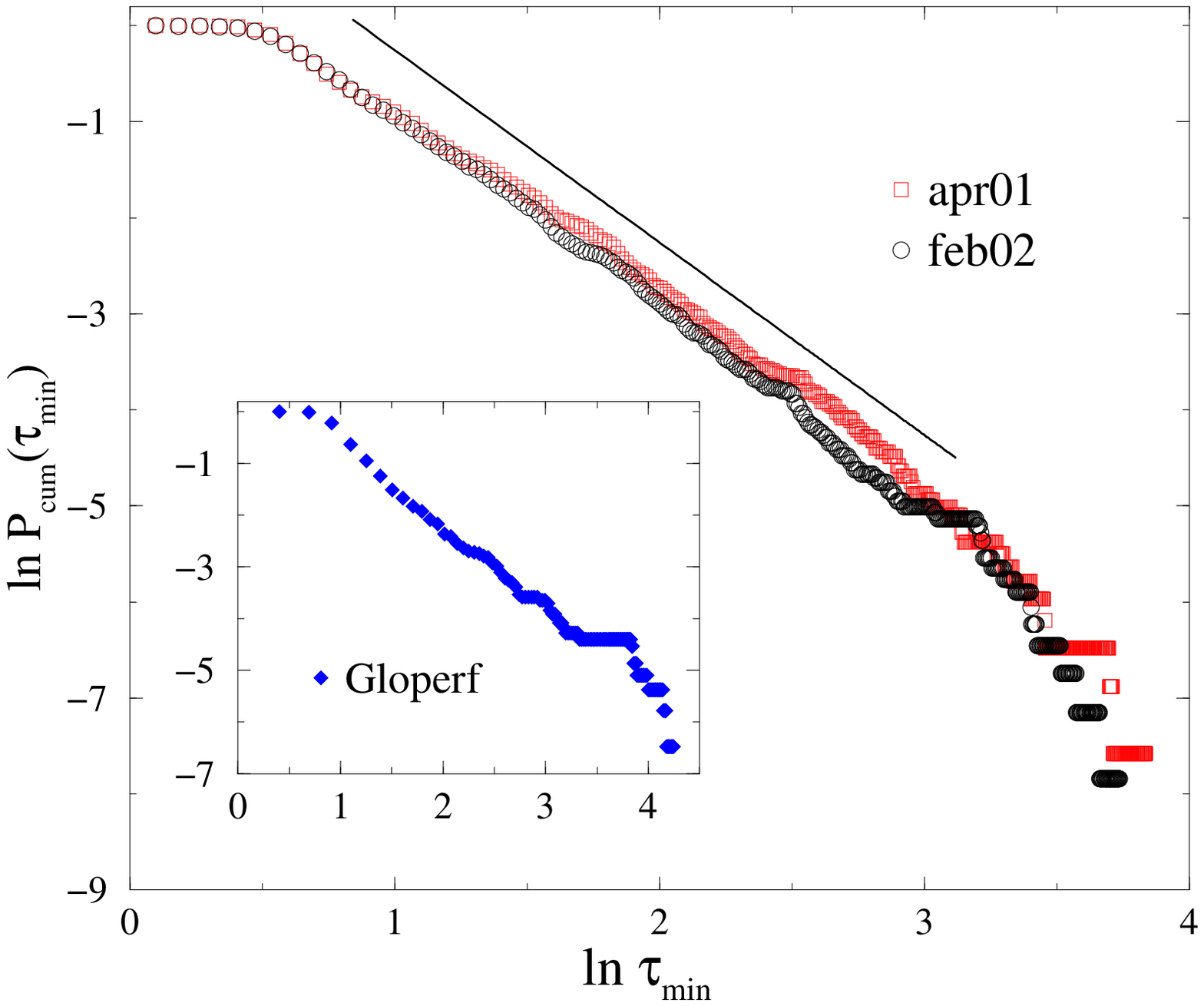}}
\centerline{\epsfxsize=8.0cm \epsfbox{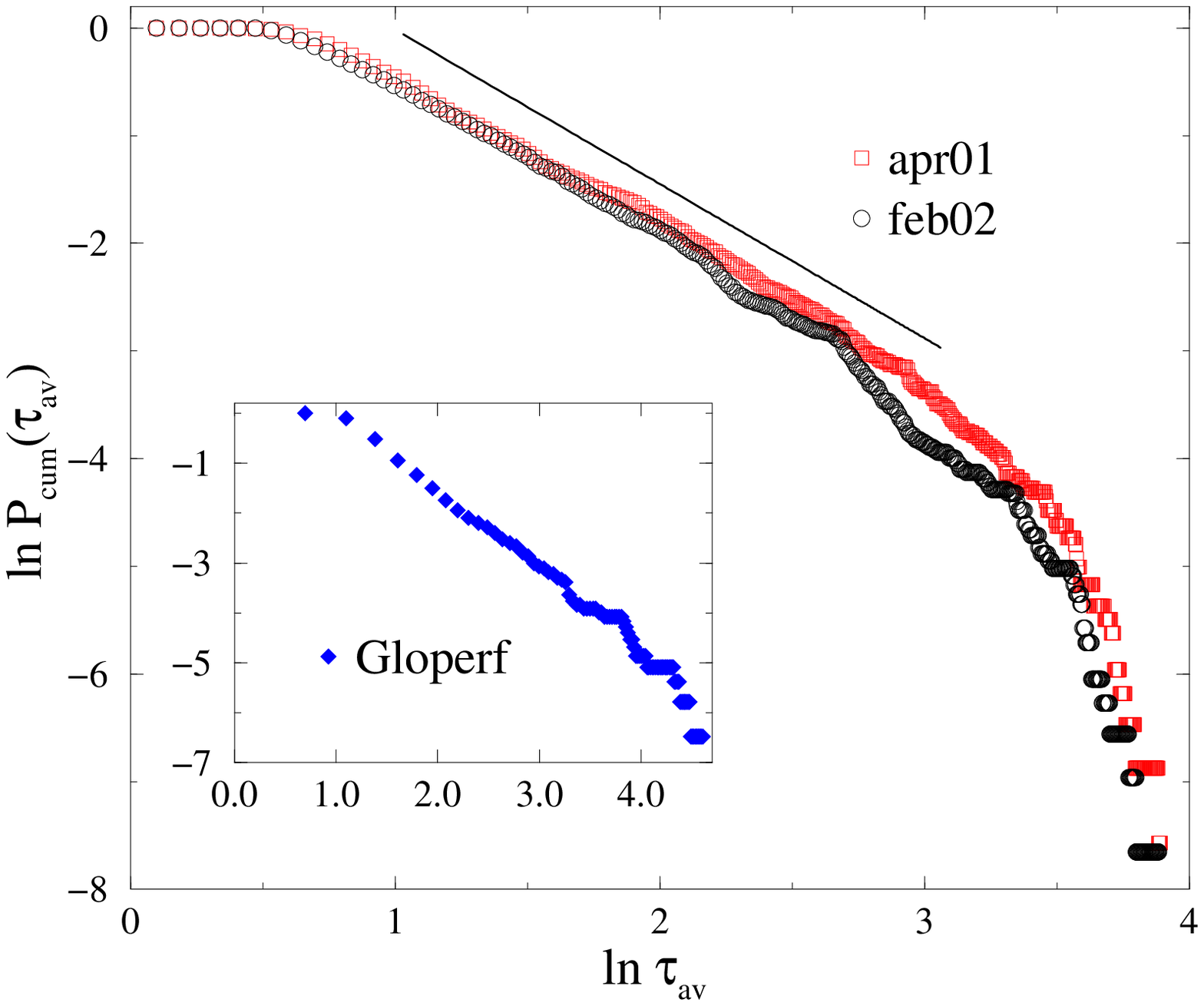}}
\caption{
{\fontfamily{cmss}\fontseries{bx}\selectfont 
Cumulative distributions}, 
of the round-trip-times normalized 
with the actual distance $d$ between host pairs.The linear behavior 
in the double logarithmic scale indicates a broad distribution
with power-law behavior. 
(a) In the case of the 
normalized minimum round-trip-times $\tau_{min}$, the slope 
of the reference line is $-2.0$. (b) In the case of the 
normalized average round-trip-times $\tau_{av}$, 
the reference line has a slope $-1.5$.
The insets of a) and b) report the distributions obtained for 
the Gloperf dataset. In both cases  we obtain  power-law behaviors 
in good agreement with those obtained for the PingER data sets (see Tab.I).\\}
\label{fig2}
\end{figure}

A more significant characterization of the end-to-end performance 
is obtained by normalizing the latency time by the geographical
distance between hosts. This defines the absolute performance metrics
$\tau_{min}=$RTT$_{min}/d$ and $\tau_{av}=$RTT$_{av}/d$ which represent 
the minimum and average latency time for unit distance, i.e. the 
inverse of the overall communication velocity (note that if we
measure $d$ in light-milliseconds $\tau_{min}$ and $\tau_{av}$ 
are actually dimensionless). These metrics allow
us to meaningfully compare the performance between pairs of hosts with 
different geographical distances.
The highly scattered plot of Fig.~\ref{fig1}, indicates that 
end-to-end performance fluctuates conspicuously  in the whole 
range of geographic distances. In particular, looking at collections
of host pairs at approximately the same geographical distance, 
we find latency times varying up to two orders of magnitude. 
\begin{figure}
\centerline{\epsfxsize=7cm \epsfbox{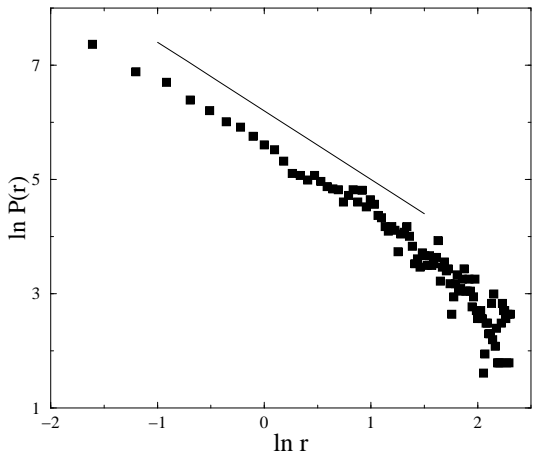}}
\caption{
{\fontfamily{cmss}\fontseries{bx}\selectfont 
Probability density $P(r)$ for the occurrence of packet
loss rate $r$}
on  beacon-target pairs transmissions.
The zero on the $x$ axis corresponds to a $1\%$ rate in packet loss.
Note that the distribution has a linear behavior in the double logarithmic
scale, indicating a power law behavior. The reference 
line has a slope $-1.2$.\\}
\label{fig3} 
\end{figure}
The best way to characterize the level of fluctuations in latency
times is represented by the probability $P(\tau_{min})$ 
and $P(\tau_{av})$ that a pair of hosts present a given $\tau_{min}$ and 
$\tau_{av}$, respectively. 
In contrast with usual exponential or 
gaussian distributions, for which there is a well defined scale, we
find that data closely follow a straight line in a double
logarithmic plot for at least one or two orders of magnitude, 
defining a power-law behavior 
$P(\tau_{min})\sim \tau_{min}^{-\alpha_{min}}$ and 
$P(\tau_{av})\sim \tau_{av}^{-\alpha_{av}}$. 
In Fig.\ref{fig2} we show  the cumulative distributions 
$P_{cum}(\tau)=\int_{\tau}^\infty P(\tau')d\tau'$
obtained from the PingER data. If the probability density distribution
is a power law  $P(\tau)\sim \tau^{-\alpha}$, the 
cumulate distribution preserves the algebraic behavior and scales as 
$P_{cum}(\tau)=\tau^{-(\alpha-1)}$. In addition, it has the advantage 
of being considerably less noisy than the original distribution.
From the behavior of Fig.\ref{fig2}, a best fit of the linear region in
the double logarithmic representation yields the scaling exponents
$\alpha_{min}\simeq 3.0$ and $\alpha_{av}\simeq 2.5$.
It is worth remarking that the presence of a truncation 
of the power law behavior for large 
values is a natural effect implicitly present in every real world 
data set and it is likely due to an incomplete statistical sampling 
of the distribution. Power-law distributions are characterized by 
scale-free properties; i.e. unbounded fluctuations and the 
absence of a meaningful characteristic length usually associated 
with the probability distribution peak.
In such a case, the mean distribution value and the
corresponding averages are poorly significant, since fluctuations
are gigantic and there are non negligible probabilities to have 
very large $\tau_{min}$ and $\tau_{av}$ compared to the average 
values in the whole system. In other words, Internet performances 
are extremely heterogeneous and it is impossible to infer 
local properties from average quantities. 

The origin of scale-free behavior is usually associated to critical 
cooperative dynamical effects. Critical and scale-free behavior 
has been observed and characterized in queueing properties at router
interfaces, probably affecting conspicuously the distribution
of $\tau_{av}$. It is, however, unclear why scale-free properties are
observed also in the distribution of $\tau_{min}$. In this case
traffic effects should be negligible, and it is well known that the
the distribution of hop counts between hosts 
has a well defined peak and no fat tails \cite{huff2}. 
On the contrary, we find that minimum latency times are distributed 
over more than two orders of magnitude.
Potentially, cables wiggliness, Internet connectivity
and hardware heterogeneities might be playing a role in the 
observed performance distribution.

It is worth remarking that a tendency to improved performance is 
observed over the two years period of data collections. 
Table I shows that the averages over all the site pairs of  
$<\tau_{min}>$ and $<\tau_{av}>$ decreases steadily,
whereas the exponents $\alpha_{min}$ and $\alpha_{av}$ increases
signalling a faster decay of the distribution tails. 
We can consider the improvement of performance as the byproduct of 
the technological drift to better lines and routers. On the other hand, 
the large fluctuations present in the Internet performance appear to 
be a stable and general feature of the statistical analysis. 
In order to have an independent check of the PingER results, we have
considered also the Gloperf data set that was used in \cite{leestep}.
We have extracted a set of parameter values for each of
650 unique site pairs in the sample and analyzed the statistics.
These results are also reported in Table I.
Although the averages depend on the specific 
characteristics of the sample (size, world region etc.) 
and differ significantly from the
PingER case, the existence of power law tails and the values
of the exponents seem to be confirmed. These exponents can thus 
be considered as one of the few and sought after reliable and 
invariant properties of the Internet \cite{paxson2}.

Finally, a  further evidence of large fluctuations in 
Internet performance is provided by the analysis of the packet loss data. 
Also in this case we are  interested in 
the probability $P(r)$ that a certain rate $r$ of packet loss occur 
on any given pair. We have analyzed the monthly average packet
loss between PingER beacon-target pairs.
In Fig.3 we report the probability $P(r)$ as a function of $r$.
The plot shows an algebraically decaying distribution that can 
be well approximated by a power-law behavior $P(r)\sim r^{-\gamma}$
with $\gamma=1.2\pm0.2$.
The slowly decaying probability of large  packet loss rate
is another signature of the very heterogeneous performance 
of the Internet.
\begin{table}
\begin{tabular}{ccccc}
Data set & $\alpha_{min}$ & $\alpha_{av}$ & $\langle \tau_{min}\rangle$ 
& $\langle \tau_{av}\rangle$ \\
\hline
April '00 & $2.7\pm0.2$ & $2.2\pm0.2$ & $3.7$ & $6.6$ \\
\hline
Feb. '01 & $2.9\pm0.2$ & $2.4\pm0.2$ & $3.6$ & $6.6$ \\
\hline
Feb. '02 & $3.0\pm0.2$ & $2.5\pm0.2$ & $3.1$ & $5.3$  \\
\hline
Gloperf &$2.7\pm0.2$ &$2.4\pm0.2$  & $5.4$ & $7.8$  \\
\hline
\end{tabular}
\caption{The table shows the improving performances along the years of 
the PingER data sample. As an independent check, we report the values 
obtained from the analysis of the data sample of the Gloperf project.}
\label{tab:1}
\end{table}
The results presented here have implications for the evaluation of
performance trends. Models for primary performance factors 
must include the high heterogeneities observed in real data.
Time and scale extrapolation for Internet performances 
can be seriously flawed by considering just the average properties.
It is likely that we will observe in the future an improvement of the
average end-to-end performance due to increased bandwidth and router
speed, but the real improvement of the Internet as a whole 
would correspond in reducing the huge statistical fluctuations  
observed nowadays. On a more theoretical side, the explanation and 
formulation of microscopic models at the origin of the scale-free
behavior of Internet performance appear challenging, to say the
least.\\

\noindent 
{\small
We thank C.~Lee for sending us the Gloperf raw data. 
We are grateful to L.~Carbone, F.~Coccetti, L.~Cottrell,
P.~Dini, Y. Moreno, R.~Pastor-Satorras and A. V\'{a}zquez 
for helpful comments and 
discussions. This work has been supported by the Internetwork Performance 
Measurement (IPM) project of Istituto Nazionale di Fisica Nucleare, and 
the European Commission - Fet Open Project COSIN IST-2001-33555.\\}

\end{document}